\documentclass{article}
\usepackage{emulateapj}

\def\asca       {{\em ASCA}\/}
\def\axaf       {{\em AXAF}\/}
\def\hst        {{\em HST}\/}
\def\einstein   {{\em Einstein}\/}
\def\rosat      {{\em ROSAT}\/}
\def\am         {$^\prime$}
\def\as         {$^{\prime\prime}$}

\def\cmsq       {~cm$^{-2}$}
\def\kms        {~km$\;$s$^{-1}$}
\def\ergs       {~erg$\;$s$^{-1}$}

\begin{document}

\submitted{Accepted for ApJ Letters, 1997 April 11}

\lefthead{ABELL 2218}
\righthead{MARKEVITCH}

\title{ABELL 2218: X-RAY LENSING, MERGER, OR BOTH?}

\author{Maxim Markevitch\altaffilmark{1}}

\affil{Harvard-Smithsonian Center for Astrophysics, 60 Garden St.,
Cambridge, MA 02138; maxim@head-cfa.harvard.edu}

\altaffiltext{1}{Also Space Research Institute, Russian Academy of
Sciences}

\begin{abstract}

Comparison of the high resolution X-ray image of A2218 obtained with the
\rosat\ HRI with the optical \hst\ image shows several interesting
correlations. The X-ray emission within a $1'$ radius core is resolved into
several components; the central dominant galaxy does not coincide with
either of them or the emission centroid. The major X-ray peak is an
elongated feature that lies between the two mass concentrations known from
the optical lensing analysis, and coincides with optical arcs at $r\simeq
20''$ from the cD galaxy.  We speculate that this may be lensed X-ray
emission, for example (but not necessarily) of the same object lensed in the
optical.  Alternatively, this feature may be a merger shock, or a gas trail
of an infalling subgroup.  Two other X-ray enhancements are close to the two
major mass concentrations.  Both lensing and a merger are likely.

Previous X-ray derivations of the A2218 mass used a $\beta$-model fit to the
data with angular resolution that blurred the features mentioned above into
a broad constant core. As the HRI data show, such a core does not exist.
Because of this, under certain assumptions and using only the improved
imaging data, the hydrostatic estimate of the projected mass within the
lensing radius can in principle be increased by a factor of $\sim 1.4$ (and
the mass within a sphere of the same radius by a factor of 2.6) compared to
previous analyses.  However, for a merging cluster, the hydrostatic analysis
is generally inapplicable. Most other lensing clusters are more distant than
A2218 and obtaining adequate X-ray images and temperature maps of them is
even more difficult. Together with the likely overestimation of mass by the
lensing analysis (as in the simulations), oversimplification of the gas
density and temperature models resulting from inadequate resolution may
account for the lensing/X-ray mass discrepancy as suggested for A2218.

\end{abstract}

\keywords{dark matter --- galaxies: clusters: individual (A2218) ---
intergalactic medium --- gravitational lensing --- X-rays: galaxies}

\section{INTRODUCTION}

A2218 at $z=0.175$ shows a strong discrepancy between cluster masses derived
from gravitational lensing in the optical (see review in Bartelmann
\& Narayan 1995) and from X-ray analysis (Miralda-Escud\'e \& Babul 1995;
Loeb \& Mao 1994). In A2218 as well as in some other clusters (e.g., Wu \&
Fang 1997 and references therein), lensing implies a projected mass within
the cylinder delineated by the observed giant arcs up to 2--3 times greater
than the value derived from X-ray data assuming hydrostatic equilibrium,
isothermality and spherical symmetry. Weak lensing results (e.g., Squires et
al.\ 1996 and references therein) are consistent with strong lensing,
although these results are still rather uncertain. Assuming that the mass
from lensing is correct, Loeb \& Mao (1994) proposed the existence of
significant nonthermal gas support in the centers of clusters, due, for
example, to gas turbulence or magnetic fields, which would compensate for
the insufficiency of thermal pressure. A simple lensing mass estimate may be
in error; cluster simulations by Bartelmann (1995) showed that the
spherically-symmetric lensing analysis should overestimate the true mass by
an average factor of 1.6 due to the presence of substructure. The remaining
discrepancy is still significant and requires further explanation.

Makino (1996) proposed a declining temperature profile to account for the
mass discrepancy in A2218; Loewenstein (1997) detected a temperature decline
in the outer ($r$ greater than a few arcminutes) region of the cluster but
found it insufficient to explain the discrepancy. A2218 has one of the
largest values of $\beta_T \equiv \mu m_p \sigma_{\rm gal}^2/ kT_e\simeq
1.6$ (where $kT_e=7.2$ keV, Mushotzky \& Loewenstein 1997, and $\sigma_{\rm
gal}=1370$\kms, Le Borgne et al.\ 1992), which is a likely indication of the
cluster's nonrelaxed state (see, e.g., simulations by Navarro et al.\ 1995).
Indeed, as Kneib et al.\ (1995) and Squires et al.\ (1996) noted, a \rosat\
X-ray image of A2218 suggests an ongoing merger of subclusters. Below we
discuss interesting details from a longer and better-positioned \rosat\ HRI
exposure, comparing it with the optical \hst\ image from Kneib et al.\
(1996) to investigate the nature of the mass discrepancy.  Although strong
temperature gradients in the inner region are likely if there is a merger,
and a large-scale temperature gradient is detected, we will not consider
their effects on the mass estimate and will limit the discussion to the
imaging data.

\section{HRI IMAGE}

The \rosat\ HRI performed four observations of A2218, two of which (with a
total exposure of 35.6 ks) are on-axis and two others offset by 12\am. We
use only the on-axis pointings in which the HRI point-spread function (PSF)
at the position of interest has a half-power diameter of 4\as\ (David et
al.\ 1996). A PSF of 20\as\ in the offset pointings is insufficient to
resolve the small scale structure discussed below. Images for the two
pointings with 5\as\ pixels were generated using the software of S. Snowden
and then co-added.  There is a bright X-ray source 12\am\ off-axis in the
HRI field of view, coincident with the bright star SAO17151. A
\rosat\ PSPC spectrum of this source is soft which is typical of stars; thus
the identification is firm. The proper motion of this star is negligible on
the relevant timescales. This star was used to correct for a $\sim 3''$
shift in the \rosat\ sky coordinates in both pointings (such an error is
well within the nominal range). Since there is only one reference source, a
possible rotational error of the HRI pointing cannot be corrected in a
similar way.  However, A. Vikhlinin (1997, private communication) infers
from the analysis of a large number of the PSPC observations that it is
likely to be negligible. It is also rather improbable that, if any
significant rotational error exists, the independent offset and rotation
combine in such a way that the X-ray star coincides with its optical
counterpart to a 3\as\ accuracy before any correction. We therefore conclude
that the corrected sky coordinates of the X-ray image are accurate to $\sim
2-3''$.

The HRI surface brightness contours are overlaid on the optical image in
Fig.\ 1. The image shows a complex multi-peaked X-ray structure in the inner
1\am--2\am\ of the cluster.  Neither of the X-ray peaks coincides with the
cD galaxy, whose J2000 coordinates from the Digitized Sky Survey image are
$\alpha=16^{\rm h} 35^{\rm m} 49^{\rm s}\!.2,\, \delta=+66^\circ 12' 45''$.
Because our reference star is approximately to the northwest of the cluster,
any uncorrected rotation around the star could not make any of the
brightness peaks coincident with the cD galaxy. Kneib et al.\ (1995), who
used one-third of this HRI exposure, and Squires et al.\ (using the same
dataset as here) reported approximate coincidence of the cD and the major
X-ray peak. Our disagreement with the former authors apparently originates
from their use of the incorrect cD galaxy coordinates (the coordinates
listed in the NASA/IPAC Extragalactic Database with the reference to J. Le
Borgne 1993 private communication, differ by 13\as\ from the position given
above). The X-ray image of Squires et al.\ appears inverted east-west with
respect to the optical image, therefore it is likely that these authors have
forced the coincidence between the X-ray peak and the cD. The
poorer-resolution PSPC and offset HRI images agree with the HRI image
presented here.

A standard $\beta$-model fit to the X-ray surface brightness profile
centered at the overall emission centroid yields $a_x=58''$ and
$\beta=0.63$, in agreement with the \einstein\ IPC (Birkinshaw \& Hughes
1994) and \rosat\ PSPC values (Squires et al.\ 1996). However, comparing the
central $100''\times 100''$ part of the image binned in 20\as\ pixels to the
two-dimensional $\beta$-model by means of a $\chi^2$ test shows that this
model is unacceptable at the 99.99\% confidence due to the structures seen
in Fig.\ 1.

\section{DISCUSSION}

\subsection{X-Ray Lensing?}

The major X-ray peak to the south of the cD has a bow-like shape and lies at
the position of the brightest optical arc. Using the \hst\ data, Kneib et
al.\ (1996) found three point-like counterparts to this arc, all four lying
along the east-to-south segment of the 20\as\ radius circle centered on the
cD (see Fig.\ 1) and being lensed images of the same object at $z=0.7$.  The
elongated X-ray structure covers about the same region.  It is quite
possible that lensed X-ray flux from the same object (or maybe other
background X-ray sources, see discussion in Refregier \& Loeb 1997)
contributes to the central X-ray structure. The flux of the brightness
excess is about 1\% of the total cluster flux. At the redshift of the
optical arc and assuming a magnification of 5--10 as for the optical arc,
this would correspond to an X-ray luminosity of the lensed source of the
order of $10^{43}$\ergs. Presently, such an interpretation of the image is
of course one of the many possibilities, since, for example, this X-ray peak
is also coincident with a bright ringed galaxy (belonging to the cluster, Le
Borgne et al.\ 1992) and a group surrounding it. The \rosat\ HRI resolution
is insufficient to determine whether this X-ray feature consists of point-
or arc-like sources (which would indicate lensing of a compact source) or is
extended, but \axaf\ instruments will be capable of resolving X-ray lensing
here. If lensing of background sources contributes any considerable X-ray
flux near the cluster center, it could modify the derived gas density
profile and hence the X-ray mass estimate.  X-ray lensing has been observed
with the HRI in another object (Chartas et al.\ 1995).

\subsection{Subcluster Merger}

Another likely explanation of the observed X-ray structure is a subcluster
merger, as was noted by previous investigators. From the lensing data, Kneib
et al.\ (1995, 1996) detected two major mass concentrations around the cD
galaxy and the second-brightest galaxy to the southeast. X-ray contours
(Fig.\ 1) show extension and perhaps a local enhancement toward the
second-brightest galaxy's concentration, while another X-ray peak lies $\sim
10''$ to the northwest of the cD. These two mass concentrations must merge.
If they are infalling head-on, then the peak ``behind'' (to the northwest
of) the cD may be offset from the cD by ram pressure. The central elongated
structure, discussed in the previous section, lies between the mass
concentrations perpendicular to the direction of the merger, and may be a
shock such as those predicted in hydrodynamic simulations (e.g., Schindler
\& M\"uller 1993; Roettiger et al.\ 1993).  Alternatively, this structure,
together with its long northeastern extension not associated with any
apparent galaxy concentration (Fig.\ 1), may be a gas trail of an infalling
group originally not belonging to either of the two main subclusters. In the
absence of a detailed gas temperature map in which merger shocks would be
apparent, construction of such merger scenarios is highly speculative.
Nevertheless, even the available X-ray image and optical lensing data
strongly suggest that there indeed is a merger underway. \axaf\ will be able
to obtain a temperature map and provide the missing information about the
merger in this cluster.

It is also possible that the absence of an X-ray peak centered on the cD is
due to partial absorption by cold material, e.g., accumulated from a past
cooling flow (White 1992). The absorbing column should be of the order of
$10^{21}$\cmsq\ and higher to be noticeable in the 0.5--2 keV image. The
\rosat\ HRI data have no energy information, and the angular resolution of
the PSPC is insufficient to test this possibility.

Note that a merger in A2218 would imply strong temperature gradients and
absence of spherical symmetry, making the Hubble constant estimates using
this cluster (e.g., Birkinshaw \& Hughes 1994; Saunders 1997) highly
uncertain (see, e.g., simulations by Roettiger, Stone \& Mushotzky 1997).

\subsection{Effect of Angular Resolution on Mass Estimate}

Miralda-Escud\'e \& Babul (1995) and Loeb \& Mao (1994) noted that the
presence of lensed images around the cD implies a mass distribution more
centrally concentrated, or, equivalently, a mass inside the 20\as\ lens
radius greater than that derived from the hydrostatic isothermal X-ray
analysis. The gas density profile they used was derived from the
$\beta$-model fits to the cluster images obtained with the \rosat\ PSPC and
\einstein\ IPC, respectively. Those fits have core radii $a_x$ around $1'$,
consistent with the HRI result given above. Essentially, it is the absence
of the radial gradient in the assumed gas density profile with $a_x=1'$ at
the lens radius of 20\as\ that gives rise to the discrepancy between the
X-ray and lensing mass measurements. However, it is now apparent from the
better-resolution HRI data that in fact 1) the centroid of the X-ray
emission is offset from the cD by about 20\as, and 2) the ``core'' is a
blend of previously unresolved brightness peaks. As is said above, the image
strongly suggests a subcluster merger and hence violent gas motions and the
likely absence of hydrostatic equilibrium, the basic condition for an X-ray
mass estimate. The gas turbulence proposed by Loeb \& Mao essentially means
the same thing.

However, in addition to this likely breakdown of the equilibrium assumption,
the limited angular resolution and the resulting oversimplification of the
gas density model also have a significant effect on the A2218 mass estimate.
If, for example, one assumes that the head-on merger scenario is correct and
the X-ray peak to the northwest of the cD has been centered on the cD in the
past, then one may expect it to retain some information about the cD
gravitational well (e.g., if the bulk motions are rapid enough, the gas
density peak may not have had enough time to disperse completely after the
removal of the cD potential). For a crude estimate, we fit a simple
symmetric $\beta$-model centered on this brightness peak, excluding other
parts of the image and not going too far off peak. The fit yields $a_x=26''$
and $\beta=0.49$. For an isothermal spherically symmetric equilibrium model,
this corresponds (see e.g., Sarazin 1988) to a mass within a 20\as\ radius
sphere 2.6 times the mass from the ``old'' $\beta$-model, and to a factor of
1.4 increase in the projected mass within the cylinder of the same radius.
This increase is due to the greater observed density gradient at the lensing
radius. This crude calculation illustrates the likely effect of the limited
angular resolution of the previous X-ray measurements on the mass at small
radii. Note that no temperature information exists for the central part of
A2218, therefore this calculation has little physical meaning and is made
solely for comparison with earlier analyses also performed under the
isothermality and equilibrium assumptions.

Most other lensing clusters are more distant than A2218 and their X-ray
images of sufficient quality are even more difficult to obtain.  Still more
problematic (practically impossible at present) is a detailed measurement of
the temperature distribution for distant clusters. Both distributions are
necessary not only for an accurate hydrostatic analysis but also to judge
its applicability. First attempts at restoring the cluster mass profiles
that take into account real temperature distributions obtained by \asca\
(e.g., Markevitch et al.\ 1996; Ikebe et al.\ 1996; Allen et al.\ 1996;
Markevitch \& Vikhlinin 1997) suggest that the mass is more centrally peaked
compared to that derived assuming isothermality. It is therefore possible
that, together with the substructure (Bartelmann 1995) and projection
effects (e.g., Daines et al.\ 1997) which tend to increase the mass derived
from lensing, inadequate modeling of gas density and temperature
distributions can account for the lensing/X-ray mass discrepancy as
suggested for A2218.

\section{SUMMARY}

The \rosat\ HRI image of A2218 reveals complex structure in the cluster core
region. One of the elongated brightness features is coincident with the
bright optical gravitational arc and its counterparts. We speculate that
some X-ray emission in this feature can also arise from lensing. Taking into
account the detailed structure of the core can significantly increase the
X-ray hydrostatic mass estimate at the lensing radius. Together with the
expected mass overestimate by the lensing analysis, this can explain the
previously reported lensing/X-ray mass discrepancy in A2218.  Moreover, the
image strongly suggests an ongoing subcluster merger, which makes the
hydrostatic mass estimate inapplicable.

\acknowledgments

I am grateful to A. Vikhlinin for discussions and help with the HRI
analysis. W. Forman, A. Loeb, and the referee, S. Mao, provided valuable
comments on the manuscript. This work was supported by NASA grant NAG5-2611.

\vspace{1cm}

{\small{\sc Fig. 1}---\rosat\ HRI X-ray contours overlaid on the optical
image. The optical $6'\times 6'$ image is from the Digitized Sky Survey,
with the central rectangular region replaced by the \hst\ image from Kneib
et al.\ (1996). The inset in the upper right corner shows a star about $12'$
off cluster center at the same angular scale.  This star was used to make a
$\sim 3''$ adjustment in the relative offset of the X-ray and optical
images, ignoring its proper motion of $1.5''$ between the epochs of the DSS
plate and the HRI observation. The HRI image is smoothed with a $\sigma=5''$
Gaussian; contours of constant surface brightness are plotted at 0.9, 2.0,
3.2, 4.0, 4.6, 5.3, 5.8, 6.4, 7.0 $\times 10^{-6}$ counts s$^{-1}$
arcsec$^{-2}$.  }

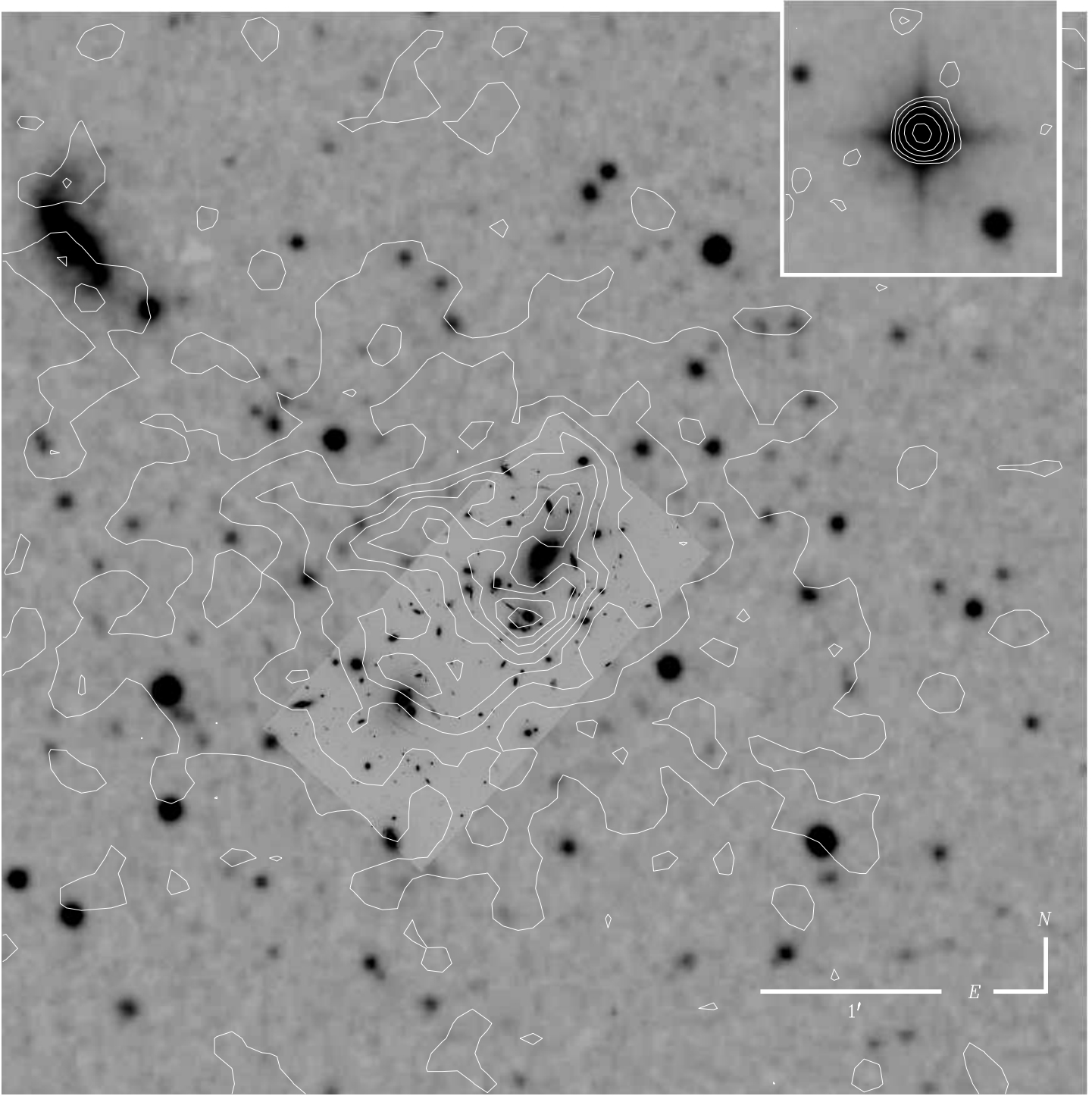
\end{document}